\documentclass[
aps,
prl,
groupedaddress,
twocolumn
]{revtex4-1}

\usepackage{graphicx}

\begin{document}


\title{Effect of the spin-orbit interaction on thermodynamic properties of liquid uranium}


\author{Minakov D.V.}
\email[]{minakovd@ihed.ras.ru}
\author{Paramonov M.A.}
\email[]{mikhail-paramon@mail.ru}
\author{Levashov P.R.}
\email[]{pasha@ihed.ras.ru}
\affiliation{%
Joint Institute for High Temperatures, Izhorskaya 13 bldg 2, Moscow 125412, Russia
}%
\affiliation{%
Moscow Institute of Physics and Technology, 9 Institutskiy per., Dolgoprudny, Moscow Region, 141700, Russia}%


\date{\today}

\begin{abstract}
We present the first quantum molecular dynamics calculation of zero-pressure isobar of solid and liquid uranium that account for spin-orbit coupling. We demonstrate that inclusion of spin-orbit interaction leads to higher degree of the thermal expansion of uranium, especially in the liquid phase. Full accounting of relativistic effects for valence electrons, particularly spin-orbital splitting of the $5f$ band, is substantial for the reproduction of the experimental density of molten uranium at the melting temperature. Influence of the spin-orbit interaction on the thermodynamic properties at high temperatures and pressures is also analyzed.
\end{abstract}


\maketitle


While influence of spin-orbit coupling (SOC) on the properties of solid uranium is intensively studied using first-principle methods~\cite{Soderlind:PRB:1990, Jones:PRB:2000, Nordstrom:PRB:2000, Soderlind:PRB:2002, Tobin:PRB:2005, Xiang:JNM:2008, Soderlind:MRS:2010, Soderlind:PRB:2012, Xie:PRB:2013} and vigorously discussed~\cite{Soderlind:PRB:2014, Xie:PRB:2016}, effect on the liquid uranium is still unclear. The reason is extreme computational complexity of such calculations for an unordered phase that requires taking into account dynamics of the atoms. Quantum molecular dynamics (QMD) simulation of uranium is essentially complex due to a large number of valence electrons, and a calculation that account for the spin-orbit (SO) interaction requires about an order of magnitude more time than a scalar-relativistic one~\cite{Smirnov:PRB:2018}.

It was shown earlier~\cite{Soderlind:PRB:1990} that the SO splitting of the $5f$ band explains the experimentally observed anomalously high room-temperature thermal expansion of light actinides, in particular, uranium, neptunium, and plutonium.
  
\begin{figure}[b]
\includegraphics[width=0.99\columnwidth]{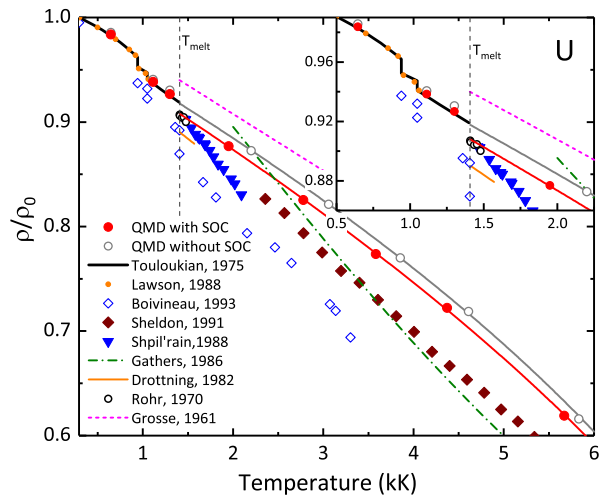}
\caption{Phase diagram of uranium in the vicinity of melting in the relative density versus temperature plane. Results of our QMD calculations of the zero-pressure isobar with account of SOC are red dots and without SOC are open grey dots. Red and grey lines are polynomial approximations of liquid state calculations. Experimental data are from~\cite{Touloukian:1975, Lawson:AC:1988, Boivineau:PPCM:1993, Sheldon:JNM:1991, Shpilrain:HT:1988, Gathers:RPP:1986, Drotning:HTHP:1982, Rohr:JPC:1970, Grosse:JACS:1961}. The inset is a closer look at the melting region.}
\label{fig:u-isobar-melting}
\end{figure}

In this Letter we demonstrate that SOC is responsible for the effect of significant increase of pressure for uranium in the vicinity of melting mostly in the liquid phase. As a consequence, lower densities along the zero isobar are predicted by QMD with SOC. Full accounting of relativistic effects for valence electrons, especially SO splitting, is substantial for the reproduction of the molten uranium density. In Fig.~\ref{fig:u-isobar-melting} we demonstrate available data of static~\cite{Touloukian:1975, Lawson:AC:1988, Shpilrain:HT:1988, Grosse:JACS:1961, Rohr:JPC:1970, Drotning:HTHP:1982} and dynamic~\cite{Gathers:RPP:1986, Sheldon:JNM:1991, Boivineau:PPCM:1993} experiments on thermal expansion of uranium in the vicinity of melting as well as results of our QMD calculations of the zero-pressure isobar with and without SOC of the valence electrons. As can be seen from the figure, accounting of SOC provides excellent agreement with data on thermal expansion of solid uranium. We can also notice discrepancy between calculations with and without SOC as temperature rises. However, the most significant difference is observed for the calculations in the liquid phase. We have found out that account of SOC increases pressure in the system by 7-8~kbar for solid uranium and by more than 10.5~kbar for liquid uranium near melting. Meanwhile, SOC makes it possible to describe with high accuracy the relative density of molten uranium at melting temperature measured in static experiments by Rohr~\textit{et al.}~\cite{Rohr:JPC:1970} and Shpil'rain~\cite{Shpilrain:HT:1988}. This value is generally accepted as a reference density of liquid uranium~\cite{CRC:2005}. However, the slope of the thermal expansion curve in liquid uranium is still a subject of debate~\cite{Iosilevskiy:JNM:2005}. As can be seen from Fig.~\ref{fig:u-isobar-melting}, experimental data on liquid uranium obtained by different authors are very contradictory. Nevertheless, the slope of our curve are in excellent agreement with measurements by Rohr~\textit{et al.}~\cite{Rohr:JPC:1970}, Grosse~\textit{et al.}~\cite{Grosse:JACS:1961}, and Drottning~\cite{Drotning:HTHP:1982}. Satisfactory agreement is also observed between our curve and the slope of the isobar dynamically measured using a pulse-heating technique by Sheldon and Mulford~\cite{Sheldon:JNM:1991}.
  
Thermodynamic properties of solid and liquid uranium are derived from QMD simulations using Vienna \textit{ab initio} simulation package (VASP)~\cite{Kresse:PRB:1993,Kresse:PRB:1994,Kresse:PRB:1996}. Calculations were carried out using finite--temperature density functional theory (FT-DFT) within the Perdew-Burke-Ernzerhof generalized gradient approximation~\cite{Perdew:PRL:1996,Perdew:PRL:1997} as implemented in VASP. The plane-wave basis set was expanded to a cutoff energy $E_{cut}$ of 500~eV. We employ a projector augmented wave (PAW)~\cite{Blochl:PR:1994,Kresse:PRB:1999} pseudopotential with 14 valence electrons. Electronic states were calculated at the Baldereschi mean--value point~\cite{Baldereschi:PRB:1973} for the liquid phase and using a 2$\times$2$\times$2 Monkhorst--Pack grid for the solid phase. The FT-DFT electronic structure calculations are performed within a collinear formulation of spin states, in other words, with spin polarization. Supercells of 108 atoms for $\alpha$-U, 128 atoms for $\gamma$-U, and 54 atoms for liquid uranium were used for simulations. The convergence with respect to the number of atoms and $\mathbf{k}$-point sampling was checked. Finite--temperature effects on the electrons were taken into account by using the Fermi--Dirac smearing. The ionic temperature was controlled by the Nos\'e-Hoover thermostat~\cite{Nose:JCP:1984}.

All QMD simulations were performed in the $NVT$ ensemble, the zero isobar was restored from calculations along isotherms in the solid phase and along isochors in the liquid phase using linear regression as described in our previous works~\cite{Minakov:AIPADV:2018, Minakov:HTHP:2020}. Since the calculated density at normal conditions slightly differs from the experimental value, it is reasonable to compare the results of calculations and measurements in the units of relative density $\rho/\rho_0$, where $\rho_0$ is a density at normal conditions (see Fig.~\ref{fig:u-isobar-melting}). In case of QMD $\rho_0=19.386$~g/cm$^3$ and 19.48~g/cm$^3$ for the calculations with and without SOC, correspondingly.

It should be mentioned, that in the conventional mode VASP performs a fully relativistic calculation for the core-electrons and treats valence electrons in a scalar relativistic approximation~\cite{Hafner:JCC:2008}. For convenience, we will further denote such calculations as ``noSOC''. Meanwhile, SOC may be also switched on for valence electrons. In VASP, the explicit implementation of SOC is based on the zeroth-order regular approximation~\cite{Lenthe:JCP:1993} and described in detail in~\cite{Steiner:PRB:2016}. We will designate such calculations as ``SOC''.

\begin{figure}
\includegraphics[width=0.9\columnwidth]{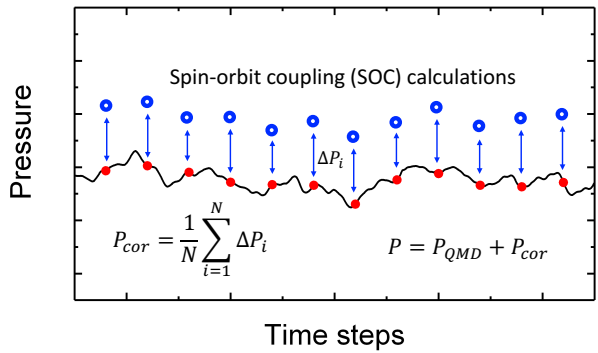}
\caption{Technique of calculation of the SOC correction to pressure for a QMD calculation. Blue dots are SOC calculations for configurations denoted as red dots.}
\label{fig:soc-calc}
\end{figure}

Since QMD simulation with SOC is tremendously time--consuming and memory demanding we have developed and applied a special correction technique. Our calculation method consists of the following steps: \textit{1) we perform a QMD simulation at a given temperature and density with spin polarization; 2) we choose several configurations on the trajectory and perform full relativistic calculations (SOC) for the chosen configurations; 3) we retrieve the pressure/energy difference between the SOC and noSOC calculations for each configuration; 4) we determine the correction to the QMD-calculated pressure/energy by averaging the differences for dozens of configurations.} The approach is schematically shown in Fig.~\ref{fig:soc-calc} for pressure evolution. We usually perform no less than 20 SOC calculations for each QMD run. We have checked statistical significance by carrying out more than 100 SOC calculations for several long simulations. Our analysis shows that SOC corrections to pressure and energy are described well by the normal distribution and averaging over 20 configurations provides a mean value that agree with averaging over 100 configurations within the standard error.

\begin{figure}
\includegraphics[width=0.99\columnwidth]{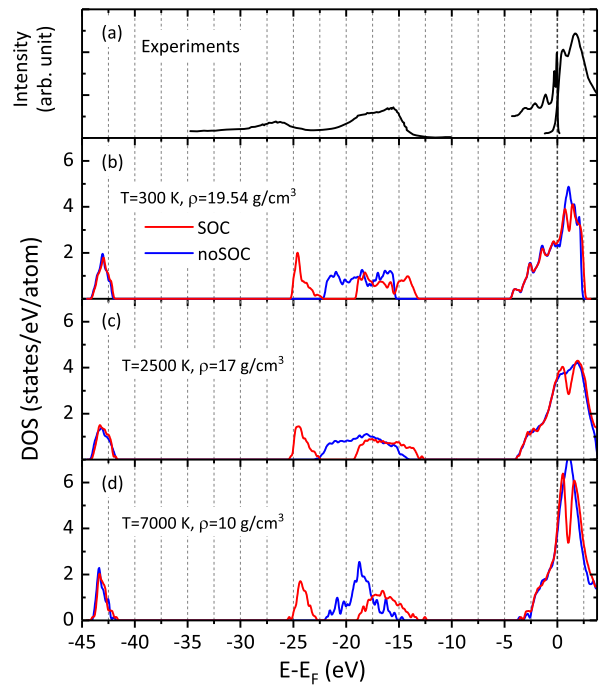}
\caption{Spectroscopic experimental data and electronic DOS from QMD simulations. (a)---experiments~\cite{Opeil:PRB:2006, Baer:PRB:1980}; (b)---QMD DOS for $\alpha$-U at $T=300$~K and $\rho=19.54$~g/cm$^3$; (c)---QMD DOS for liquid U at $T=2500$~K and $\rho=17$~g/cm$^3$; (d)---QMD DOS for liquid U at $T=7000$~K and $\rho=10$~g/cm$^3$. All energies with respect to the Fermi level.}
\label{fig:u-dos-qmd}
\end{figure}

In order to understand the reason for significant effect of the SO interaction on the thermal expansion of solid and liquid uranium we analyze the electronic density of states (DOS). DOS restored from QMD simulations for solid uranium at $T=300$~K and $\rho=19.54$~g/cm$^3$ is shown in Fig.~\ref{fig:u-dos-qmd}(b), and for liquid uranium at $T=2500$ and $T=7000$~K in Fig.~\ref{fig:u-dos-qmd}(c) and Fig.~\ref{fig:u-dos-qmd}(d), respectively. Smooth curves were obtained by averaging over QMD snapshots and applying a Gaussian smearing of 0.1~eV to the band energies. Experimental data~\cite{Opeil:PRB:2006, Baer:PRB:1980} are also shown in Fig.~\ref{fig:u-dos-qmd}(a). X-ray photoemission spectroscopy (XPS)~\cite{Fuggle:JPMP:1974,McLean:PRB:1982,Opeil:PRB:2006} revealed spin-orbit splitting of the U $6p$ core-states (9.5~eV). Bremsstrahlung isochromat spectroscopy (BIS) allowed to investigate the electronic DOS above the Fermi level and revealed the SO splitting of $5f$ states with a separation of 1.15~eV~\cite{Baer:PRB:1980}. As can be seen from the figure, both effects of splitting of $6p$ and $5f$ states, as well as the valence band spectrum near the Fermi level obtained by ultraviolet photoemission spectrosopy (UPS)~\cite{Opeil:PRB:2006}, are described very well by our QMD-SOC calculations. The splitting remains at higher temperatures and lower densities.

We investigate the influence of SOC during the thermal excitation of valence electrons more precisely using static DFT calculations. In this case we calculate the pressure difference between SOC and noSOC calculations for 3 different crystal lattices: $\alpha$-U, $\gamma$-U (bcc), and fcc. Fcc is a close-packed structure with high coordination number (12), which is close to our estimate of the first coordination number of uranium liquid near melting (14). Densities are chosen to correspond to $\alpha$, $\gamma$, and liquid phases of U from QMD calculations, respectively. Static calculations were performed with a higher energy convergence criterion for the electronic loop (10$^{-7}$~eV) and finer $\mathbf{k}$-point grid (15$\times$15$\times$15).
\begin{figure}
\includegraphics[width=0.99\columnwidth]{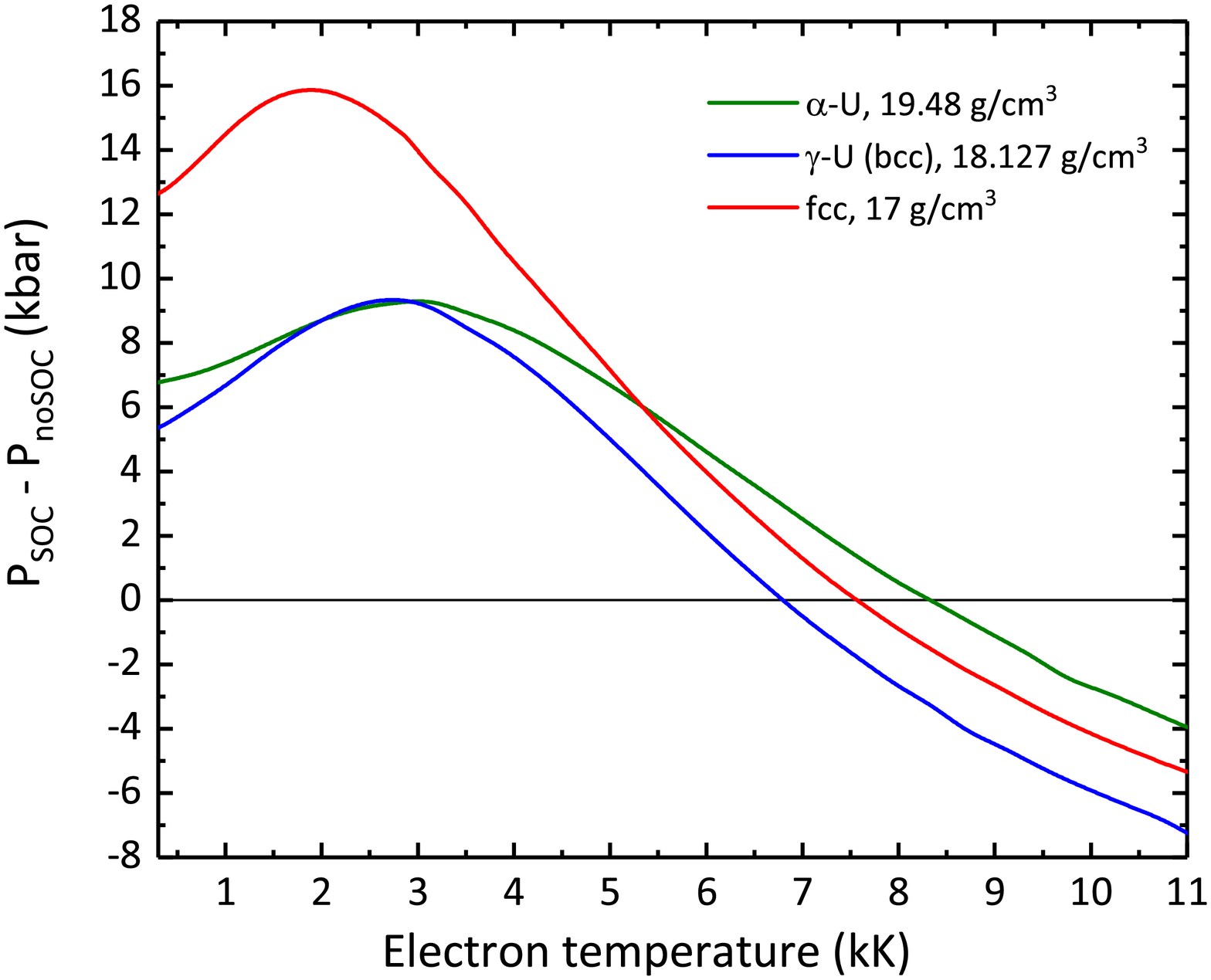}
\caption{Pressure difference between SOC and noSOC static FT-DFT calculations as a function of electron temperature for different lattices and densities.}
\label{fig:pdif-crystal}
\end{figure}
\begin{figure}
\includegraphics[width=0.99\columnwidth]{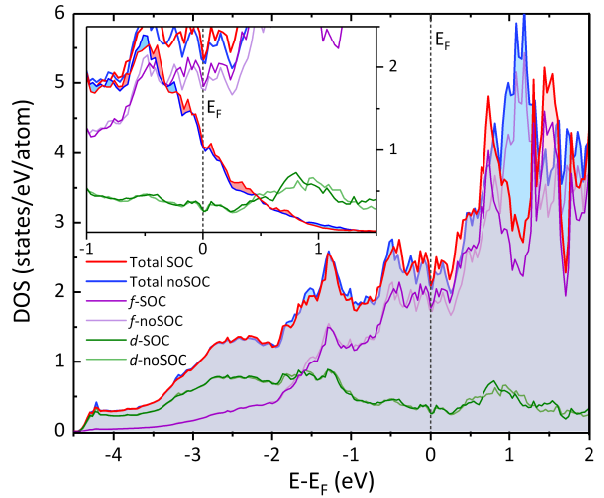}
\caption{Total and $6d$ and $5f$ partial DOS in $\alpha$-U at $\rho=19.48$~g/cm$^3$ from static FT-DFT calculation. All energies with respect to the Fermi level. The inset shows occupied DOS in the vicinity of the Fermi level from SOC and noSOC calculations as thin red and blue lines, respectively.}
\label{fig:pdos}
\end{figure}
Figure~\ref{fig:pdif-crystal} shows that the pressure difference between SOC and noSOC calculations has an explicit maximum in the range of 2--3~kK for all considered lattices. We note here that the peak is significantly higher for the fcc lattice. We also present the total and partial electronic DOS for $\alpha$-U from static DFT calculation in Fig.~\ref{fig:pdos} to study the effect of thermal excitation of valence electrons in more detail. It can be clearly seen that the SO splitting of $5f$ states is responsible for the redistribution of electronic states from higher energies closer to the region of the Fermi level. We demonstrate the occupied states at 3~kK in the inset in Fig.~\ref{fig:pdos}. For convenience we fill the area where the occupied DOS from the SOC calculation is above (below) the one from the noSOC calculation with red (blue). Apparently, the SO splitting leads to a higher electron occupancy in the vicinity of the Fermi level and above, which in turn results in higher electronic pressure in a certain temperature range compared to noSOC calculation. 
At higher temperatures the occupancies of the excited electrons for energies above 1~eV is lower in SOC calculations than in noSOC ones due to the gap between the shoulders of the split $5f$ band, so the opposite effect of negative pressure difference is observed.

As can be seen from Fig.~\ref{fig:pdif-crystal} the effect of the spin-orbit interaction on pressure depends not only upon temperature but structure as well, and it may be stronger for denser-packed structures. This cumulative effect seems to be able to explain noticeably more intensive influence of SOC on thermal expansion of liquid uranium than that of solid in the vicinity of melting, as well as further diminishing of the discrepancy between SOC and noSOC calculations at higher temperatures.

\begin{figure}
\includegraphics[width=0.99\columnwidth]{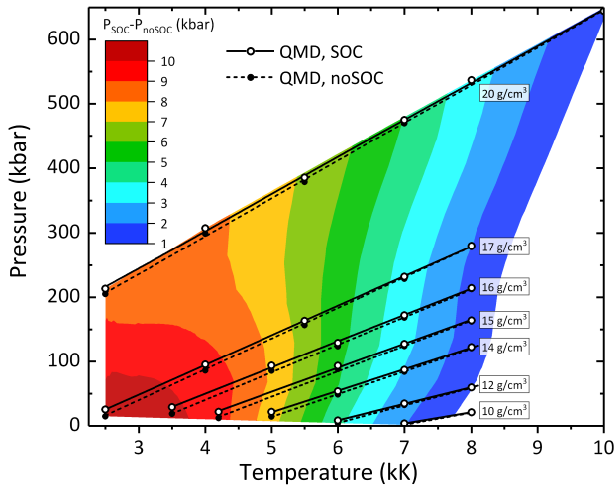}
\caption{QMD-calculated pressure with respect to temperature along isochors for liquid uranium. Open black dots are results of QMD simulations with SOC corrections to pressure, solid black dots are results without corrections. Solid and dashed lines are linear fits for isochors from SOC and noSOC calculations, respectively. Color contour map indicates smoothed values of SOC corrections to pressure.}
\label{fig:pdif-surface}
\end{figure}

In order to provide a broader picture of the influence of the SO interaction on the thermodynamic properties of liquid uranium at higher temperature and pressure, we present the dependence of the pressure difference between QMD SOC and noSOC calculations upon temperature along the set of isochors in Fig.~\ref{fig:pdif-surface}. The color contour map illustrates a complex effect of pressure and temperature. Nevertheless, it is consistent with our previous static DFT analysis in general. Interestingly, there is an area of the maximum impact of the SOC near the melting line at low pressures. On the other hand for higher temperatures influence of the SOC becomes stronger as pressure rises.

In conclusion, this study demonstrates, for the first time, that relativistic effects have substantial impact on thermodynamic properties of liquid uranium, especially on density at the melting point. The reason is the SO splitting of the $5f$ band and the thermal excitation of valence electrons. We have analyzed the influence of the SOC on the thermodynamic properties of liquid uranium at high temperatures and pressures and revealed parameters where the effect of SOC can be neglected. Finally, this study become possible due to the technique of QMD simulation with account of SOC that was developed and successfully applied in this Letter.

We appreciate sincerely Prof Igor Iosilevskiy for the motivation for this work.
We acknowledge the JIHT RAS Supercomputer Center, the Joint Supercomputer Center of the Russian Academy of Sciences, and the Shared Resource Centre ``Far Eastern Computing Resource'' IACP FEB RAS for for providing computing time.
\begin{acknowledgments}
The work is supported by the Russian Science Foundation (grant No.\,18-79-00346).

\end{acknowledgments}


%

\end{document}